\documentclass[aps,twocolumn,superscriptaddress,showpacs,showkeys]{revtex4}
\usepackage{epsfig}
\usepackage{graphicx}
\usepackage{bm}
\usepackage{natbib}
\usepackage{dcolumn}
\usepackage{amsmath}
\usepackage{rotating}
\usepackage{multirow}

\begin{document}

\title{Renormalization flows in complex networks}

\author{Filippo Radicchi\footnote{Correspondence should be addressed to FR.
Electronic address: f.radicchi@gmail.com}}
\affiliation{Complex Systems and Networks
 Lagrange Laboratory (CNLL), ISI Foundation, Turin, Italy}
\author{Alain Barrat}
\affiliation{CPT (CNRS UMR 6207), Luminy Case 907, F-13288 Marseille Cedex 9, France}
\affiliation{Complex Systems and Networks Lagrange Laboratory (CNLL), ISI Foundation, Turin, Italy}
\author{Santo Fortunato}
\affiliation{Complex Systems and Networks Lagrange Laboratory (CNLL), ISI Foundation, Turin, Italy}
\author{Jos\'e J. Ramasco}
\affiliation{Complex Systems and Networks Lagrange Laboratory (CNLL), ISI Foundation, Turin, Italy}

\begin{abstract}
Complex networks have acquired a great popularity in
recent years, since the graph representation of many natural,
social and technological systems is often very helpful
to characterize and model their phenomenology. Additionally, the mathematical tools of
statistical physics have proven to be particularly
suitable for studying and understanding complex
networks. Nevertheless, an important obstacle to this theoretical
approach is still represented by the difficulties to draw parallelisms
between network science and more traditional aspects of 
statistical physics. In this paper, we explore the relation
between complex networks and a well known topic of statistical physics:
renormalization. A general method to analyze renormalization flows of
complex networks is introduced. The method can be applied to study any
suitable renormalization transformation. Finite-size scaling can be
performed on computer-generated networks in order to classify them in
universality classes. We also present applications of the method on real networks.
\end{abstract}

\maketitle

\section{Introduction}
\label{sec:intro}
Many real systems in nature, society and technology can be represented
as complex networks
\cite{albertbarabasirev,newmanreview,dorogovtsevmendesbook,pastorvespignanibook,boccaletti06,barratbook}.
Independently of their natural, social or technological origin, most
networks share common topological features, like the ``small-world'' property \cite{watts98}
and a strong topological
heterogeneity. The small-world property expresses the fact that the average distance between nodes, as
defined in the graph-theoretical sense, is small with respect to the
number of nodes, and typically grows only logarithmically with it.
Networks are topologically heterogeneous in that
the distributions of the number of neighbors (degree) of a node are broad, typically spanning 
several orders of magnitude, with tails that can often be described by power laws 
(hence the name ``scale-free networks'' \cite{barabasi99}). 

While ``scale-free-ness'' implies the absence of a characteristic
scale for the degree of a node, it is not a priori
clear how this can be related to the notion of self-similarity, 
often studied in statistical physics, and also typically related
to the occurrence of power laws. In this context, several 
recent works have focused on defining and studying
the concept of self-similarity for
networks. The notion of self-similarity is related to a
renormalization transformation, properly adapted to graphs,
introduced by Song {\it et al.}~\cite{song05}. The renormalization
procedure is analogous to standard length-scale transformations, used in
classical systems~\cite{stanley71,cardy96}, and can be simply
performed by using a box covering technique interpreted in a
graph-theoretical sense. The analysis of this transformation in real
networks~\cite{song05} has revealed that some of them, such as the
WWW, social, metabolic and protein-protein interaction networks,
appear to be self-similar while others, like the Internet, do not.
Self-similarity here means that the statistical features of a network remain
unchanged after the application of the renormalization
transformation. Many successive papers have focused on this subject,
performing the same analysis on several networks, introducing new
box-covering techniques and trying to explain the topological
differences between self-similar and non-self-similar
networks~\cite{yook05,song06,goh06,song07,kim07a,kim07b,kim07c} (for a
recent review on this topic see~\cite{rozenfeld08}).

In this context, the analysis of renormalization flows of complex
networks~\cite{radicchi08} represents a new perspective to study
block transformations in graphs. Differently from all former studies,
the study of Ref.~\cite{radicchi08} is not devoted to observe the effect of
a single transformation, but to analyze the renormalization flow
produced by repeated iterations of the transformation. Starting
from an initial network, the iteration of the renormalization
procedure allows to explore the space of network configurations
just as standard renormalization is used to
explore the phase space of classical systems in statistical
physics~\cite{stanley71,cardy96}. For these reasons, the analysis of
renormalization flows of complex networks represents not only an
important theoretical step towards the understanding of
block-transformations in graphs, but also a further attempt to link
traditional statistical physics and network science.

In this paper, we substantially extend the analysis presented
in~\cite{radicchi08}. We perform a numerical study of renormalization
flows for several computer-generated and real networks. The numerical
method is applied to different renormalization transformations. For a
particular class of transformations, we find that the renormalization
flow leads non-self-similar networks to a condensation transition,
where a few nodes attract a large fraction of all links. The main
result of the paper lies in the robustness of the scaling rules
governing the renormalization flow of a network: independently of the
transformation, the renormalization flow of non-self-similar networks
is characterized by the same set of scaling exponents, which identify a
unique universality class. In contrast, the renormalization flow of
self-similar networks allows to classify these networks in different
universality classes, characterized by a set of different scaling
exponents.  

The paper is organized in the following way. In
section~\ref{sec:renorm}, we describe the standard technique used in
order to renormalize a network and define the
renormalization flow of a graph. We then start with the analysis of
renormalization flows of different networks. In the case of
computer-generated graphs, we distinguish the behavior of
non-self-similar (section~\ref{sec:nssn}) and self-similar
(section~\ref{sec:ssn}) networks. Section~\ref{sec:real} is devoted to
the analysis of the renormalization flows of real complex
networks. Finally, in section~\ref{sec:concl} we summarize and comment
the results.

\section{Renormalizing complex networks}
\label{sec:renorm}

\begin{figure}
\begin{center}
\includegraphics[width=0.45\textwidth]{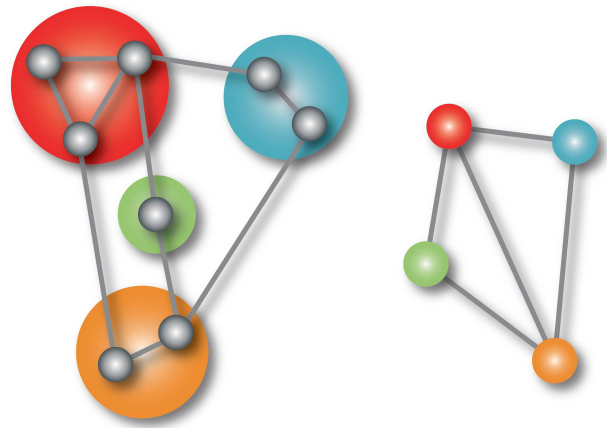}
\end{center}
\caption{Renormalization procedure applied to a simple
  graph. (left) The original network is divided into boxes and the
  renormalized graph (right) is generated according to this tiling.}
\label{fig:method}
\end{figure}

Differently from classical systems, graphs are not embedded in
Euclidean space. As a consequence, standard length-scale
transformations cannot be performed on networks since measures of
length have a meaning only in a graph-theoretical sense: the length of a
path is given by the number of edges which compose the path; the
distance between two nodes is given by the length of the (or one of
the) shortest path(s) connecting the two nodes. Based on this
metrics, Song {\it et al}.~\cite{song05} proposed an original
technique for renormalizing networks (see
Fig.~\ref{fig:method}). Given the length of the transformation
$\ell_B$, their method is given by the following steps:
\begin{itemize}
\item{Tile the network with the minimal number of boxes $N_B$; each
  box should satisfy the condition that all pairs of nodes within the
  box have distance less than $\ell_B$.}
\item{Replace each box with all nodes and mutual edges inside with a supernode.}
\item{Construct the renormalized network composed of all supernodes:
  two supernodes are connected if in the original network there is at
  least one link connecting nodes of their corresponding boxes.}
\end{itemize}

The former recipe represents a transformation $R_{\ell_B}$ applicable
to any unweighted and undirected network leading to the generation of
a new unweighted and undirected network, the ``renormalized'' version
of the original one. In principle, there are many ways to tile a
network and therefore the transformation $R_{\ell_B}$ is not
invertible. Moreover, finding the best coverage of a network (i.e.,
the one which minimizes the number of boxes $N_B$) is computationally
expensive: up to now, the best algorithm introduced in this context is
the greedy coloring algorithm~\cite{song07} (GCA), a greedy technique
inspired by the mapping of the problem of tiling a network to
node-coloring, a well known problem in graph
theory~\cite{bollobas}. An analogous technique, leading to a
qualitatively and quantitatively similar transformation $R_{r_B}$, is
random burning (RB)~\cite{goh06}. In RB boxes are spheres of
radius $r_B$ centered at some seed nodes, so that the maximal distance
between any two nodes within a box does not exceed $2r_B$. Nodes in boxes defined
through the transformation $R_{r_B}$ satisfy the condition defining boxes 
of the transformation $R_{\ell_B}$, for $\ell_B=2r_B+1$. However, the search for
minimal box coverage is much more effective for the GCA than for RB, 
and this may occasionally yield different results, as we shall see.

The strict meaning of self-similarity is that any part of an object, however small, looks like the whole~\cite{mandelbrot}. Similarly,  complex networks are self-similar if their statistical properties are invariant under a proper renormalization transformation. Song {\it et al.}~\cite{song05} have shown that the degree distribution of  several real networks remains unchanged if a few iterations of the renormalization transformation are performed. Moreover, when this  feature is verified, the number of boxes $N_B$ needed to tile the network for a given value of the length parameter $\ell_B$ decreases as a power law as $\ell_B$ increases:
\begin{equation}
N_B\left( \ell_B \right) \sim \ell_B^{-d_B}\;\;.
\label{eq:self}
\end{equation}
The exponent $d_B$ is called, in analogy with classical systems, the fractal exponent of the network~\cite{mandelbrot}. This property has been verified for several real networks in various studies~\cite{song05,yook05,goh06}. On the other hand, not all
real networks are self-similar, i.e. Eq.(\ref{eq:self}) and the invariance of the degree distribution do not hold
for them. For consistency, these networks are called non-self-similar.


In contrast to former studies which mostly dealt with a single
step of renormalization, we are interested here in
analyzing renormalization {\em flows} of complex networks, i.e. the outcome of repeated
iterations of the renormalization procedure described above. 
Starting from a graph $G_0$, with
$N_0$ nodes and $E_0$ edges, we indicate as $G_t$ (with $N_t$ nodes
and $E_t$ edges) the network obtained after $t$ iterations of the
transformation $R$:
\begin{equation}
\begin{array}{l}
G_1=R\left(G_0\right) \;,\; 
G_2=R\left(G_1\right)=R^2\left(G_0\right) \;,\; \ldots 
\\
\ldots \;,\; G_t=R\left(G_{t-1}\right)=\ldots=R^t\left(G_0\right)\;.
\end{array}
\label{eq:flux}
\end{equation}  
Note that in Eq.(\ref{eq:flux}) we have suppressed the subscript
$\ell_B$ (or $r_B$) for clarity of notation.  In our analysis, we
follow the flow by considering several observables. We mainly focus on
the variables
\begin{equation}
\kappa_t = K_t/\left(N_t-1\right) \;\;,
\label{eq:kappa}
\end{equation} 
where $K_t$ is the largest degree of the graph $G_t$, and 
\begin{equation}
\eta_t=E_t/\left(N_t-1\right) \;\;,
\label{eq:eta}
\end{equation} 
which is basically the average degree of the graph $G_t$ divided by
two. $\kappa_t$ and $\eta_t$ can assume non-trivial values in any
graph, excluding trees (for which $\eta_t=1$, $\forall t$). We monitor also the fluctuations of the
variable $\kappa_t$ along the flow by measuring the susceptibility
\begin{equation}
\chi_t = N_0 \left( \langle \kappa_t^2 \rangle - \langle \kappa_t
\rangle^2\right) \;\;,
\label{eq:chi} 
\end{equation}
where $\langle \cdot \rangle$ denotes averages
taken over different realizations of the covering algorithm.
Moreover, we consider other quantities like the
average clustering coefficient $C_t$~\cite{watts98}. All these observables
are monitored as a function of the relative network size
$x_t=N_t/N_0$, which is a natural way of following the renormalization
flow of the variables under study.

\section{Computer-generated networks}
\label{sec:artificial}

We first consider artificial networks. In the case of
computer-generated networks, it is in fact possible to control the
size $N_0$ of the initial graph $G_0$ and to perform the well known
finite-size scaling analysis for the renormalization flow. For every
computer-generated graph and every transformation $R$, we find that
the observable $\kappa_t$ obeys a relation of the type
\begin{equation}
\kappa_t = F\left[ \left(x_t - x^*\right) N_0^{1/\nu} \right]\;\;,
\label{eq:scaling}
\end{equation}
where $F\left(\cdot\right)$ is a suitable function depending on the
starting network and the particular transformation used. Analogous
scaling relations hold for the other observables ($\eta_t$ and $C_t$)
we considered. The susceptibility $\chi_t$ needs an additional
exponent $\gamma$ since it obeys a relation of the type: 
$\chi_t = N_0^{\gamma/\nu} G\left[ \left(x_t - x^*\right) N_0^{1/\nu} \right]$,
with $G\left(\cdot\right)$ a suitable scaling function. In general,
the scaling exponent $\nu$ does not depend on the particular
transformation $R$ used to renormalize the network, but depends on the
starting network $G_0$: we always obtain $\nu=2$ for any
non-self-similar network (Sec.~\ref{sec:nssn}) and values of $\nu$
depending on the initial network in the case of self-similar graphs
(Sec.~\ref{sec:ssn}). On the other hand, we obtain $x^*=0$ in all
cases, except for the particular transformations obtained for $r_B=1$
and $\ell_B=2$ on non-self-similar networks
(Sec.~\ref{subsec:nssn1}). In the next sections, we show our numerical
results, obtained from the analysis of renormalization flows of
computer-generated networks, distinguishing between the various cases.
All values of $\nu$ and $x^*$ are listed in Table~\ref{table}. 
We emphasize the importance of the fact that the exponent $\nu$
is able to classify artificial networks in different universality
classes.

\subsection{Non-self-similar networks}
\label{sec:nssn}

We consider several computer-generated networks for which
Eq.(\ref{eq:self}) does not hold. Eq.(\ref{eq:scaling}) is able to
describe the renormalization flows of any of these network models. The
scaling exponent $\nu=2$ identifies a single universality class for all these models.
The only difference is given by the finite value
of $x^*>0$ obtained when renormalization is performed with $\ell_B=2$
or $r_B=1$ (Sec.~\ref{subsec:nssn1}). Instead, for $\ell_B>2$ and
$r_B>1$ we always obtain $x^*=0$ (Sec.~\ref{subsec:nssn2}).

\subsubsection{$r_B=1$, $\ell_B=2$.}
\label{subsec:nssn1}

For $r_B=1$ or $\ell_B=2$, the transformation $R$ has a
particular behavior. In the case of GCA and $\ell_B=2$, at each stage
of the renormalization flow, the boxes in which the network is tiled
have the peculiarity to be fully connected subgraphs or {\it cliques}~\cite{derenyi05}. 
In the case of renormalization
performed with RB and $r_B=1$, spheres are compact subgraphs composed
only of neighbors of the selected seed nodes. In both cases, at each
stage of the renormalization flow, the contraction of the network is
much slower if compared with the same transformations run for higher
values of $\ell_B$ or $r_B$.  

\begin{figure*}[!ht]
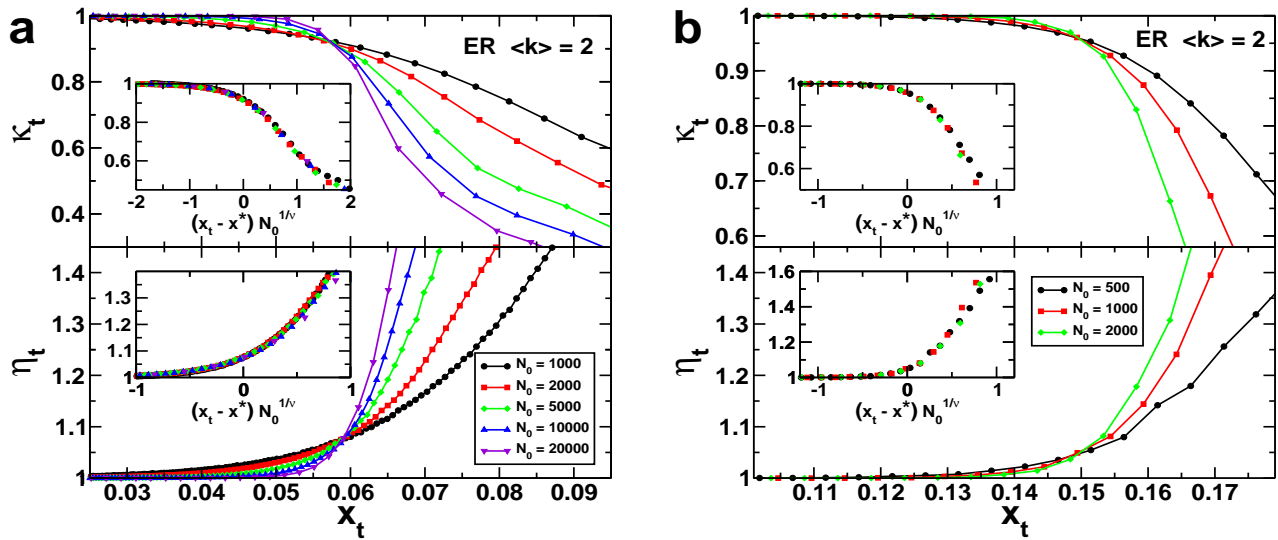

\begin{center}
\includegraphics[width=0.45\textwidth, height=0.3\textheight]{fig2a}
\qquad
\includegraphics[width=0.45\textwidth, height=0.3\textheight]{fig2b}
\end{center}
\caption{Study of renormalization flows on the ER model
  with $\langle k \rangle =2$. The box covering has been performed by
  using RB with $r_B=1$ (a) and GCA with $\ell_B=2$ (b).  The figures
  display $\kappa_t$ (a,b top) and $\eta_t$ (a,b bottom) as a function
  of the relative network size $x_t$.  The insets display the scaling
  function of the variable $(x_t-x^*)N_0^{1/\nu}$ for $\kappa_t$ and
  $\eta_t$. Here the scaling exponent $\nu=2$ in both cases. Note that the flow of the renormalization procedure goes from larger (right on the x-axis) to smaller values (left) of $x_t$.}
\label{fig:er2}
\end{figure*}

\begin{figure*}[!ht]
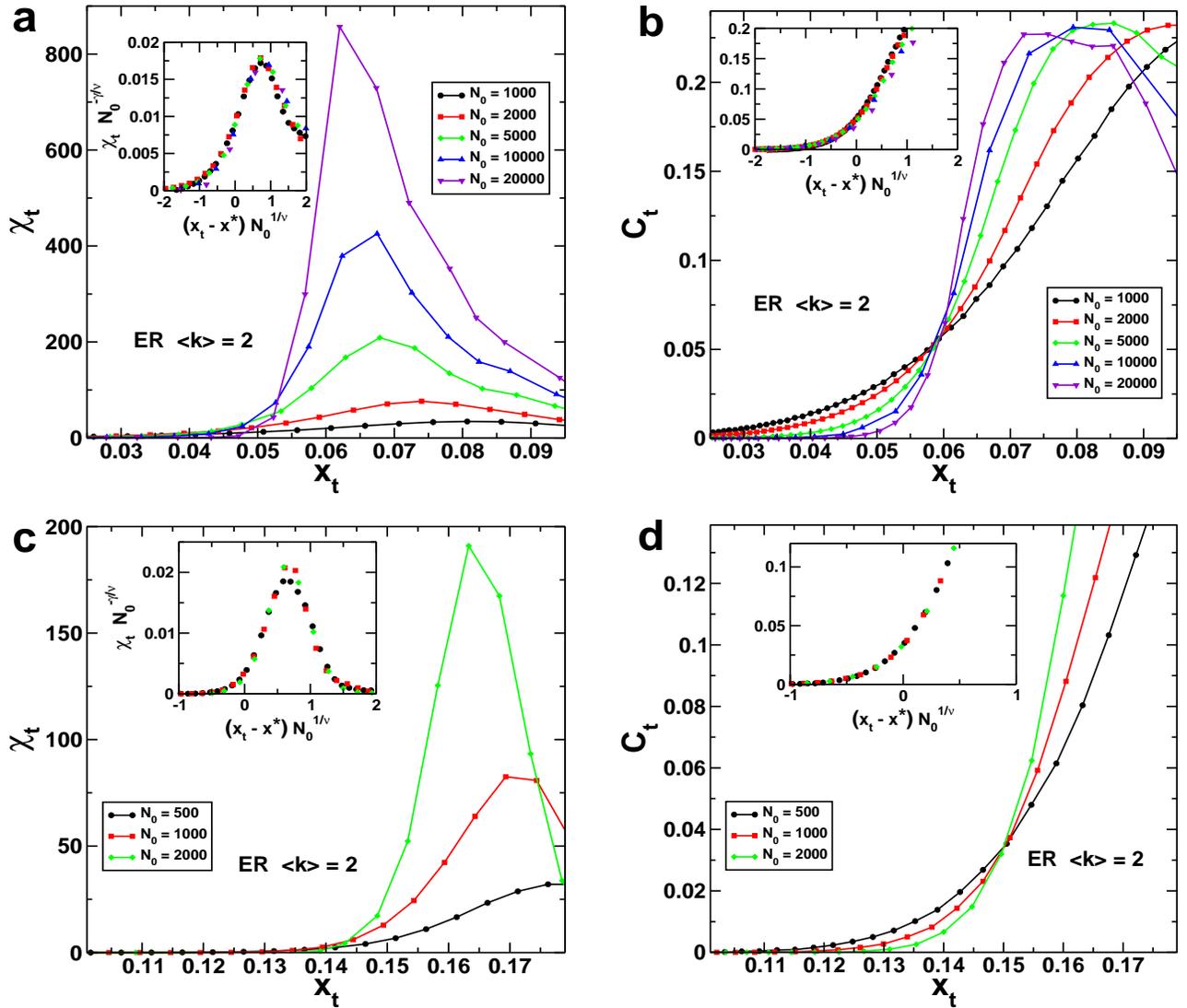

\begin{center}
\includegraphics[width=0.45\textwidth, height=0.3\textheight]{fig3a}
\qquad
\includegraphics[width=0.45\textwidth, height=0.3\textheight]{fig3b}
\vskip .3cm 
\includegraphics[width=0.45\textwidth, height=0.3\textheight]{fig3c}
\qquad
\includegraphics[width=0.45\textwidth, height=0.3\textheight]{fig3d}
\end{center}
\caption{Study of renormalization flows on ER model
  with $\langle k \rangle =2$. The box covering has been performed by
  using RB with $r_B=1$ (a,b) and GCA with $\ell_B=2$ (c,d).  The
  figures display the susceptibility $\chi_t$ (a,c) and the average
  clustering coefficient $C_t$ (b,d) as a function of the relative
  network size $x_t$.  The insets display the scaling function of the
  variable $(x_t-x^*)N_0^{1/\nu}$ for $\chi_t$ and $C_t$. Here the
  scaling exponent $\nu=2$ and the susceptibility exponent
  $\gamma=\nu$ in all cases.}
\label{fig:er22}
\end{figure*}

In Fig.~\ref{fig:er2}, we show some
numerical results obtained following the renormalization flow of the
Erd\"os-R\'enyi (ER)~\cite{erdos59} model with average degree $\langle
k \rangle =2$. For both algorithms used for renormalizing
the networks, we clearly see a point of intersection between all the
curves occurring at a particular value $x^*>0$.
Interestingly, the same values of $\nu$ and $x^*$ hold also for the
susceptibility $\chi_t$ and the average clustering coefficient
$C_t$. Numerical results for both quantities and their relative
scaling are reported in Fig.~\ref{fig:er22}.

\begin{figure*}[!ht]
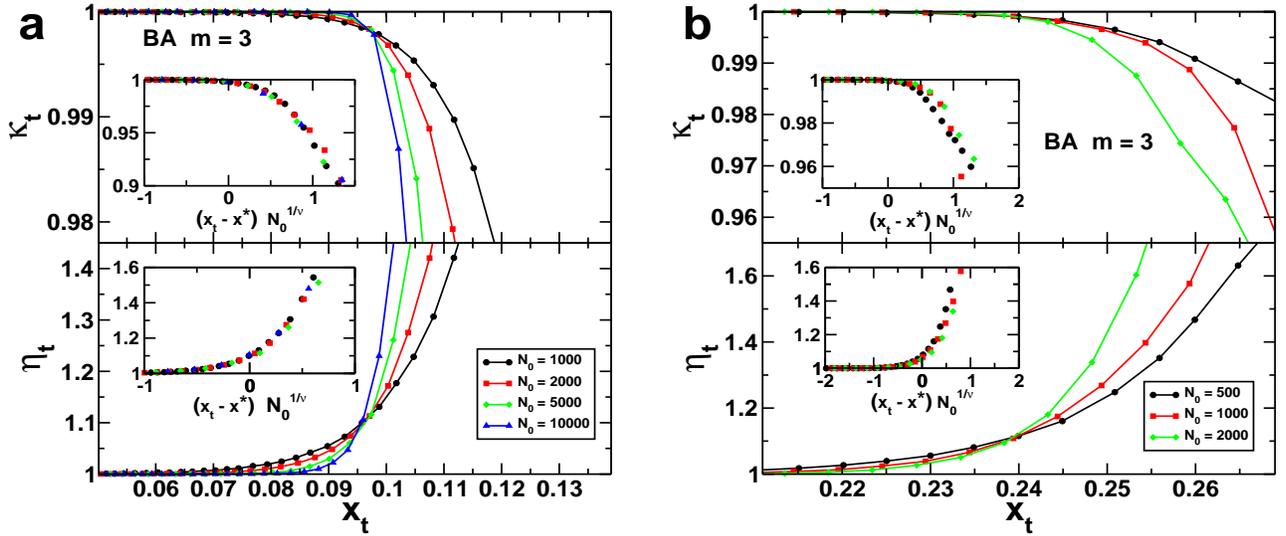

\begin{center}
\includegraphics[width=0.45\textwidth, height=0.3\textheight]{fig4a}
\qquad
\includegraphics[width=0.45\textwidth, height=0.3\textheight]{fig4b}
\end{center}
\caption{Study of renormalization flows on the BA model
  with $2m=\langle k \rangle =6$ ($m$ indicates the number of
  connections introduced by each node during the construction of the
  BA model). The box covering has been performed by using RB with
  $r_B=1$ (a) and GCA with $\ell_B=2$ (b).  The figures display the
  variables $\kappa_t$ (a,b top) and $\eta_t$ (a,b bottom) as a
  function of the relative network size $x_t$.  The insets display the
  scaling function of the variable $(x_t-x^*)N_0^{1/\nu}$ for
  $\kappa_t$ and $\eta_t$. Here the scaling exponent $\nu=2$ in both
  cases.}
\label{fig:ba}
\end{figure*}

The same behavior is observed for all the non-self-similar networks that we have studied. To mention a few, we performed numerical simulations also on the
Barab\'asi-Albert (BA) model~\cite{barabasi99} and its generalization
given by scale-free networks generated with linear preferential
attachment~\cite{dorogovtsev00}. We report in Fig.~\ref{fig:ba} the numerical results
obtained for the BA model: the quantities $\kappa_t$ and $\eta_t$ are
shown as a function of the renormalization flow's variable
$x_t$. Again, a clear crossing point $x^*>0$ can be seen in this
case. More importantly, both variables $\kappa_t$ and $\eta_t$ obey
Eq.(\ref{eq:scaling}) with $\nu=2$.

\begin{figure*}[!ht]
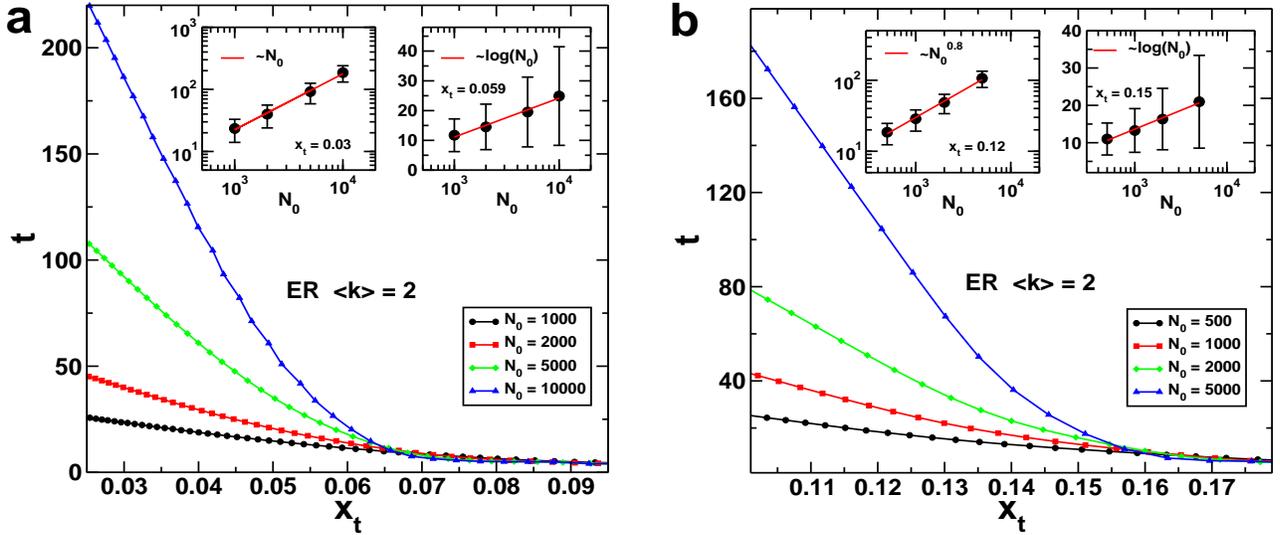

\begin{center}
\includegraphics[width=0.45\textwidth, height=0.3\textheight]{fig5a} \qquad
\includegraphics[width=0.45\textwidth, height=0.3\textheight]{fig5b}
\end{center}
\caption{Study of renormalization flows on the ER model
  with $\langle k \rangle =2$. The box covering has been performed by
  using RB with $r_B=1$ (a) and GCA with $\ell_B=2$ (b). The figures
  display the number of iterations $t$ as a function of the relative
  network size $x_t$. The fixed point [i.e., $x_t=x^*=0.059$ (RB), $0.015$ (GCA)] is reached in
  a number of renormalization steps growing logarithmically 
  with the initial system size $N_0$ (see the insets on the right in each figure). In
  contrast, the number of stages needed to go out from the fixed-point
  (i.e., to reach a given $x_t<x^*$) grows as a power of $N_0$ (see
  the insets on the left in each figure).}
\label{fig:time}
\end{figure*}

\begin{figure*}[!ht]
\begin{center}
\includegraphics[width=0.45\textwidth, height=0.3\textheight]{fig6a}
\qquad
\includegraphics[width=0.45\textwidth, height=0.3\textheight]{fig6b}
\vskip .3cm 
\includegraphics[width=0.45\textwidth, height=0.3\textheight]{fig6c}
\qquad
\includegraphics[width=0.45\textwidth, height=0.3\textheight]{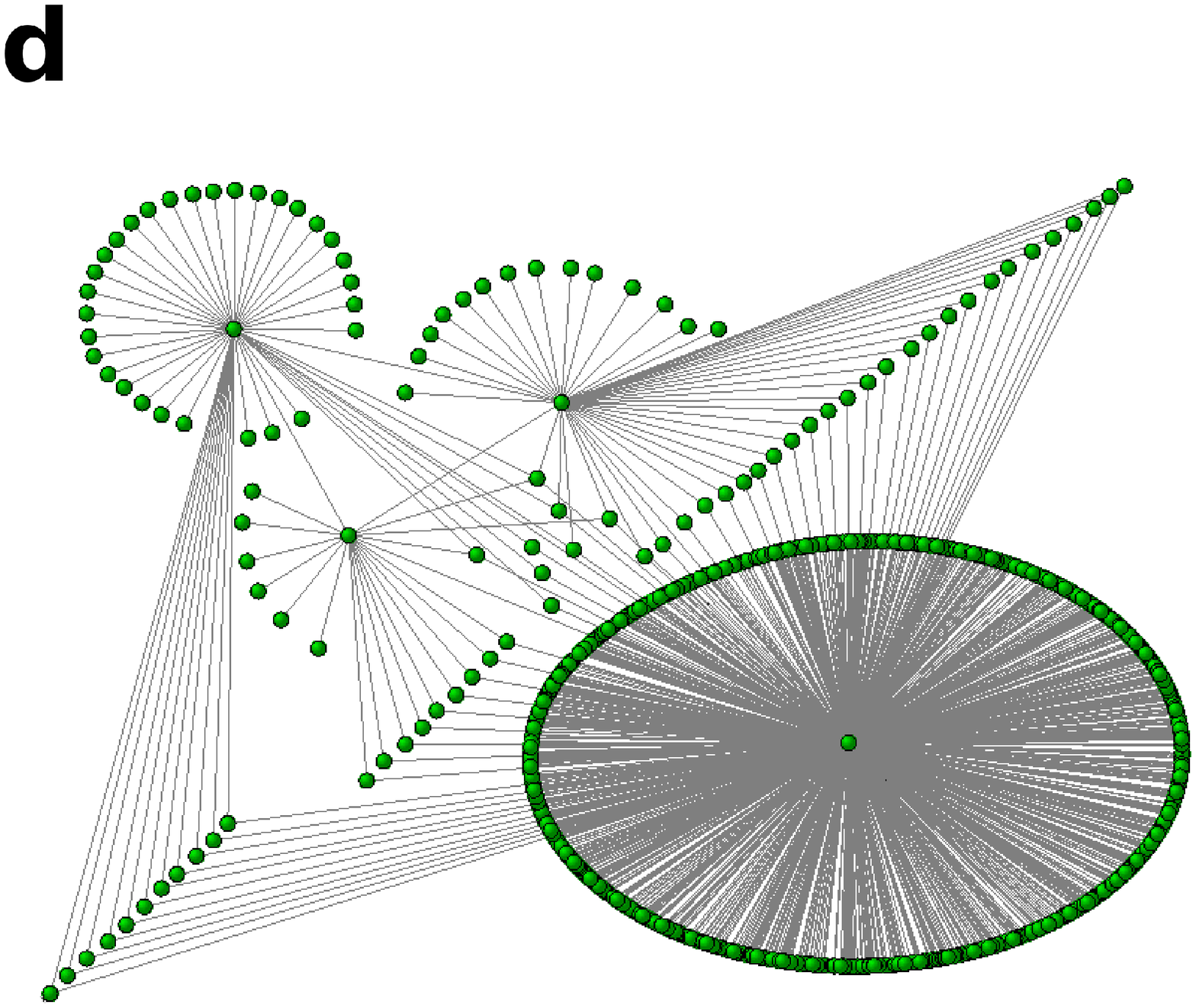}
\end{center}
\caption{Statistical properties of the fixed point in
  the case of computer-generated networks. Renormalization has been
  performed by using GCA with $\ell_B=2$. Initial network sizes are:
  $N_0 = 10^6$ for the ER model, $N_0 = 10^6$ for the BA model, $N_0 = 10^6$
  for the Watts-Strogatz (WS) model (with ratio of rewired links $p=0.01$),
  and $N_0 = 156251$ for the fractal model (FM), with ratio of added
  connections $p=0.05$. Dashed lines have slopes $-1.5$ in (a), $-1.2$
  in (b) and $-1$ in (c). (d) The graphical representation of the
  fixed point has been obtained by starting from an ER model with
  $N_0 = 30000$ and $\langle k \rangle = 2$.}
\label{fig:fixpoint}
\end{figure*}

The existence of a non-vanishing $x^*$ is a peculiarity of the
renormalization obtained for $\ell_B=2$ and $r_B=1$: $x^*>0$ implies
the existence of a special stable fixed point, which holds in the
limit of infinite network size. The fixed point is reached in a number
of iterations which scales logarithmically with the initial size of
the network, while the number of renormalization stages needed to
reach any $x_t < x^*$ diverges almost linearly with the initial system
size (see Fig.~\ref{fig:time}). Interestingly, the fixed point
statistically corresponds to the same topological structure,
independently of the topology of the initial network (see
Fig.~\ref{fig:fixpoint}). This particular fixed point is a graph where
a few nodes attract a large fraction of all links [i.e.,
  $\kappa_t\left(x^*\right)\approx 1$]; such hub nodes have
degrees which are distributed according to a power law (see
Fig.~\ref{fig:fixpoint}a). Moreover, the network obtained at the
fixed point is composed of nodes with clustering coefficient ($C$) and
average degree of the neighbors ($k_{nn}$) which decrease as a power
law as the degree of the node increases (see Figs.~\ref{fig:fixpoint}b
and~\ref{fig:fixpoint}c, respectively). Fig.~\ref{fig:fixpoint}d
is a graphical representation of the graph obtained at $x^*$
when starting from an ER network with $N_0=30000$. The presence of
many star-like structures gives an explanation of the results described
above (Fig.~\ref{fig:time}): for $r_B=1$, the center of the first
chosen box will be with high probability a low-degree node (``leaf'' in the figure), so the box will include 
only a low-degree node and the attached hub, and the
other low-degree nodes (the other leaves of the star) will need one
box each. This is why $N_t$ decreases very
slowly. Renormalization steps with 'small' boxes ($r_B=1$ or
$\ell_B=2$) make it therefore 'difficult' or 'slow' to modify appreciably
such structures.

\subsubsection{$r_B>1$, $\ell_B>2$.}
\label{subsec:nssn2}

\begin{figure*}[!ht]
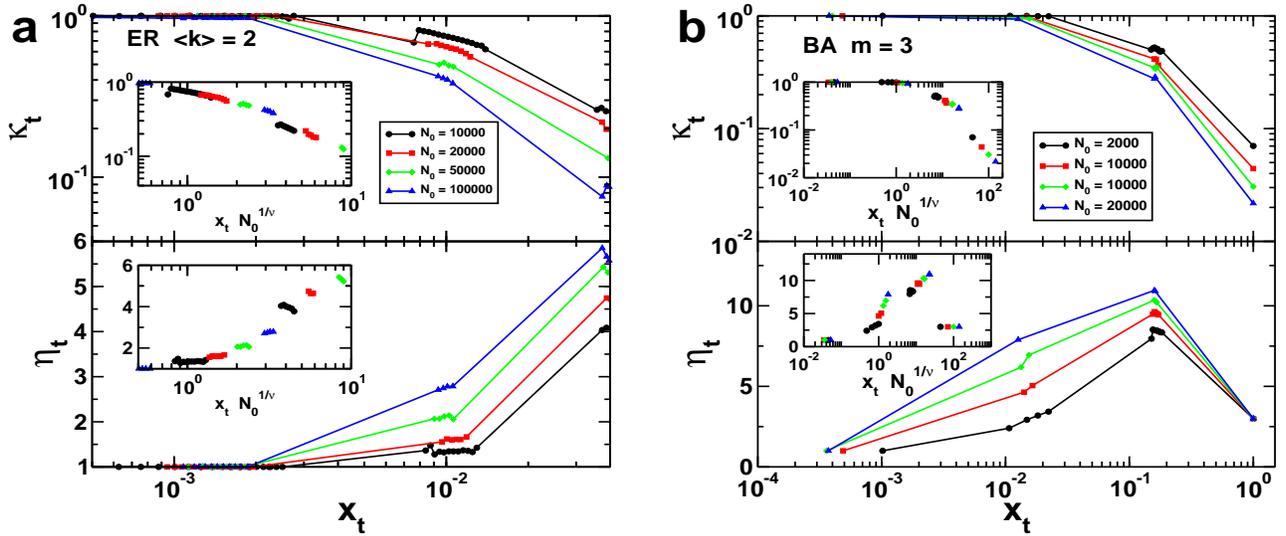

\begin{center}
\includegraphics[width=0.45\textwidth, height=0.3\textheight]{fig7a}
\qquad
\includegraphics[width=0.45\textwidth, height=0.3\textheight]{fig7b}
\end{center}
\caption{Study of renormalization flows on
  non-self-similar artificial graphs. Renormalization has been
  performed by using GCA with $\ell_B=3$. The figures display the
  variables $\kappa_t$ (a,b top) and $\eta_t$ (a,b bottom) as a
  function of the relative network size $x_t$, for the renormalization
  flow of an ER model with $\langle k \rangle =2$ (a) and a BA model
  with $2m= \langle k \rangle =6$ (b). The insets display the scaling
  function of the variable $(x_t-x^*)N_0^{1/\nu}$ for $\kappa_t$ and
  $\eta_t$. Here the scaling exponent is $\nu=2$ in both cases.}
\label{fig:lb3}
\end{figure*}

Renormalization flows with $\ell_B>2$ and $r_B>1$ have been discussed
in \cite{radicchi08}, and we present some additional results in
Fig.~\ref{fig:lb3}. The renormalization flows of non-self-similar
computer-generated networks still obey Eq.(\ref{eq:scaling}) with
$\nu=2$, and the  main difference with respect to the particular cases
$\ell_B=2$ and $r_B=1$ consists in the value of $x^*$, which is now
$x^*=0$. As usual in statistical physics, however, the precise
value of the threshold is less relevant than the value of the
exponents describing the flow. In this respect, the robustness
of the value $\nu=2$ strikingly shows that all non-self-similar
artificial networks can be classified as belonging to a unique
universality class.

\subsection{Self-similar networks}
\label{sec:ssn}

Let us now consider computer-generated models which satisfy
Eq.(\ref{eq:self}). For all these models, we find that
Eq.(\ref{eq:scaling}) still holds. In strong contrast
with non-self-similar networks, the value of the scaling exponent
$\nu$ now depends on the particular network analyzed and on the specific
renormalization transformation. Moreover, 
each self-similar model is a fixed point of the
renormalization flow since the statistical properties of the network
are unchanged if iterated renormalization transformations are applied
to the network.

As a prototype of computer-generated self-similar network, we
consider the Fractal Model (FM) introduced by Song et
al.~\cite{song06}.  The FM is self-similar by design, as it is
obtained by inverting the renormalization procedure. At each step, a
node turns into a star, with a central hub and several nodes with
degree one.  Nodes of different stars can then be connected in two
ways: with probability $e$ one connects the hubs with each other, with probability $1-e$ a non-hub of a star is connected to a
non-hub of the other. The resulting network is a tree with
power law degree distribution, the exponent of which depends on the
probability $e$.

\begin{figure*}[!ht]
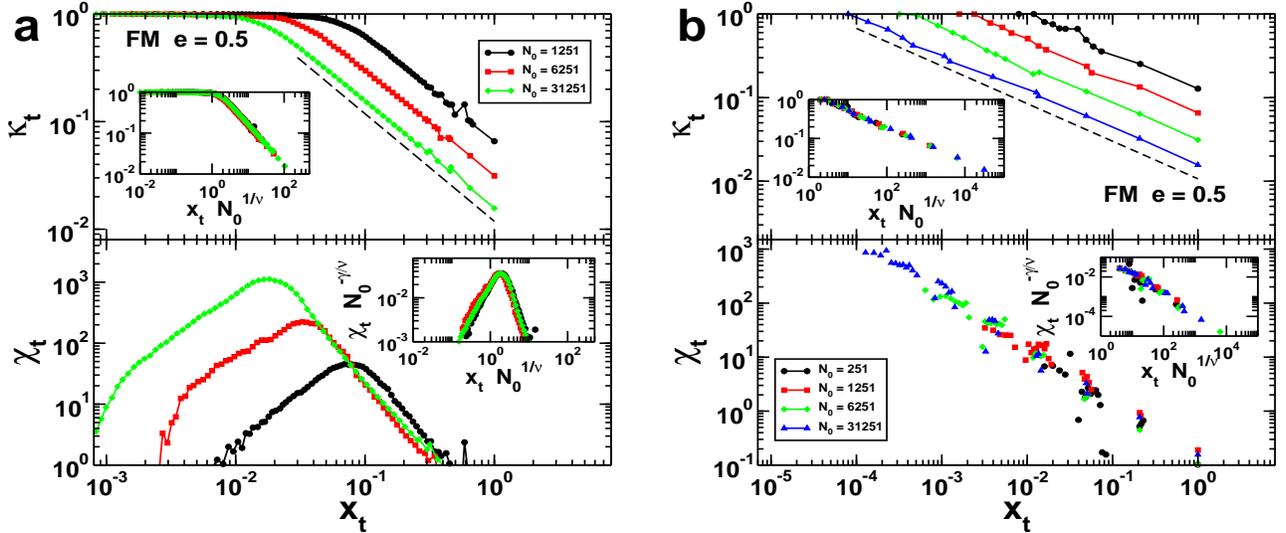

\begin{center}
\includegraphics[width=0.45\textwidth, height=0.3\textheight]{fig8a}
\qquad
\includegraphics[width=0.45\textwidth, height=0.3\textheight]{fig8b}
\end{center}
\caption{Study of renormalization flows on self-similar
  artificial graphs. The box covering has been performed by using RB
  with $r_B=1$ (a) and GCA with $\ell_B=3$ (b). The graph is an FM
  network with $e=0.5$, where $e$ is the probability for hub-hub
  attraction~\cite{song06}.  The figures display $\kappa_t$ (a, b,
  top), and $\chi_t$ (a, b, bottom) as a function of the relative
  network size $x_t$.  The scaling function of the variable
  $(x_t-x^*)N_0^{1/\nu}$ for $\kappa_t$ and $\chi_t$ is displayed in
  the insets.  We find that the two box covering techniques yield
  different exponent values: $\nu=2.2$ (RB) and $\nu=1$ (GCA). The
  dashed lines indicate the predicted behavior of the scaling
  function. In (a) the exponent of the power law decay for the scaling
  function is $-1$, independently of the exponent $\beta$ of the
  degree distribution of the initial graph; in (b) instead the scaling
  function decays with an exponent $-(\beta-2)/(\beta-1)=-0.45$.  We
  still find $\gamma=\nu$ for both transformations.}
\label{fig:fm}
\end{figure*}

In the case of the FM network it is possible to derive the scaling
exponent $\nu$, by inverting the construction procedure of the
graph. In this way one recovers graphs with identical structure at
each renormalization step and one can predict how $\kappa_t$, for
instance, varies as the flow progresses.  Since we are interested in
renormalizing the graph, our process is the time-reverse of the growth
described in~\cite{song06}, and is characterized by the following
relations
\begin{equation}
\begin{split}
N_{t-1}=n\,N_t, \\
k_{t-1}=s\,k_t, \\
\beta=1+\frac{\log\,n}{\log\,s},\\
\end{split}
\label{eq:fm1}
\end{equation}
where $n$ and $s$ are time-independent constants determining the value
of the degree distribution exponent $\beta$ of the network. Here $N_t$
and $k_t$ are the number of nodes and a characteristic degree of the
network at step $t$ of the renormalization; we choose the maximum
degree $K_t$. The initial network has size $N_0$ and shrinks due to
box-covering transformations. In this case, for the variable
$\kappa_t$ one obtains
\begin{equation}
\begin{split}
\kappa_t \sim \frac{K_t}{N_t}= \frac{K_0}{N_0}\left( \frac{s}{n}
\right)^{-t}= \frac{K_0}{N_0}\left(
\frac{N_t}{N_0}\right)^{-\frac{\beta-2}{\beta-1}}\\ =\frac{K_0}{N_0}x_t^{-\frac{\beta-2}{\beta-1}}\sim
(N_0\,x_t)^{-\frac{\beta-2}{\beta-1}},
\end{split}
\label{eq:fm2}
\end{equation}
where we used $s=n^{1/(\beta-1)}$, $N_t/N_0=n^{-t}$ and $K_0 \sim
N_0^{1/(\beta-1)}$, derived from Eqs.~(\ref{eq:fm1}). We see that the
scaling exponent $\nu=1$ is obtained for any value of the exponent
$\beta$. From Eq.(\ref{eq:fm2}) we actually get the full shape of the
scaling function, that is a power law: our numerical calculations
confirm this prediction (see Fig.~\ref{fig:fm}b).  We remark that this
holds only because one has used precisely the type of transformation
that inverts the growth process of the fractal network. This
amounts to applying the GCA with $\ell_B=3$, as we did Fig.~\ref{fig:fm}b. 

If we consider instead the renormalization
procedure defined by RB with $r_B=1$ (or by GCA with $\ell_B=2$), the
centers of the boxes will be mostly low degree nodes, as discussed
above. Hubs are thus
included in boxes only as neighbors of low degree nodes and, as a
consequence, the supernode corresponding to a box with a large hub
inside will have a degree which is essentially the same as the degree
of the hub before renormalization. It is therefore reasonable to
assume that $K_t\sim K_0$, and we get
\begin{equation}
\kappa_t \sim \frac{K_t}{N_t}\sim \frac{K_0}{N_t} \sim
\frac{N_0^{1/(\beta-1)}}{N_t}=(N_0^\frac{\beta-2}{\beta-1}x_t)^{-1}
\label{eq:fm3}
\end{equation}
which is again a scaling function of the variable $N_0^{1/\nu}x_t$,
with $\nu=(\beta-1)/(\beta-2)$, as we found numerically (see
Fig.~\ref{fig:fm}a).

Qualitatively similar numerical results can be shown also for other
self-similar models of networks: unperturbed Watts-Strogatz (WS)
model~\cite{watts98} (i.e., one-dimensional lattice), hierarchical
model~\cite{ravasz03} and the Apollonian (AP) network
model~\cite{andrade05} (see Table~\ref{table}).

\begin{table}
\begin{tabular}{| c | c | c | c | c |}
\hline
Type & Network & $R$ & $\nu$ & $x^*$
\\
\hline
\hline
\multirow{8}{*}{Non-self-similar} & \multirow{4}{*}{ER $\langle k \rangle =2$ } &  $r_B=1$ & $2.0(1)$ & $0.059(1)$
\\
& & $\ell_B=2$ & $2.0(1)$ & $0.15(1)$
\\
& & $r_B=2$ & $2.0(1)$ & $0$
\\
& & $\ell_B=3$ & $2.0(1)$ & $0$
\\
\cline{2-5}
& \multirow{4}{*}{BA $m=3$ } &  $r_B=1$ & $2.0(1)$ & $0.098(2)$
\\
& & $\ell_B=2$ & $2.0(1)$ & $0.245(5)$
\\
& & $r_B=2$ & $2.0(1)$ & $0$
\\
& & $\ell_B=3$ & $2.0(1)$ & $0$
\\
\cline{1-5}
\multirow{7}{*}{Self-similar} & \multirow{3}{*}{WS $\langle k \rangle =4$ } &  $r_B=1$ & $1.0(1)$ & $0$
\\
& & $\ell_B=2$ & $1.0(1)$ & $0$
\\
& & $\ell_B=3$ & $1.0(1)$ & $0$
\\
\cline{2-5}
& \multirow{3}{*}{FM $e=0.5$ } &  $r_B=1$ & $2.2(1)$ & $0$
\\
& & $\ell_B=2$ & $2.2(1)$ & $0$
\\
& & $\ell_B=3$ & $1.0(1)$ & $0$
\\
\cline{2-5}
& \multirow{2}{*}{AP} &  $\ell_B=2$ & $4.8(2)$ & $0$
\\
& &  $\ell_B=3$ & $1.0(1)$ & $0$
\\
\cline{1-5}
\multirow{6}{*}{{\scriptsize Perturbed self-similar}} & \multirow{2}{*}{{\scriptsize WS $\langle k \rangle =4$}} &  $r_B=1$ & $2.0(1)$ & $0.004(2)$
\\
& & $\ell_B=3$ & $2.0(1)$ & $0$
\\
\cline{2-5}
& \multirow{2}{*}{{\scriptsize FM $e=0.5$}} &  $r_B=1$ & $2.1(1)$ & $0.118(2)$
\\
& & $\ell_B=3$ & $2.0(1)$ & $0$
\\
\cline{2-5}
& \multirow{3}{*}{AP} &  $r_B=1$ & $2.0(1)$ & $0.045(2)$
\\
&  &  $\ell_B=2$ & $2.0(1)$ & $0.05(1)$
\\
&  &  $\ell_B=3$ & $2.0(1)$ & $0$
\\
\cline{1-5}
\end{tabular}
\caption{We list the values of the scaling exponent $\nu$ and of the fixed point threshold $x^*$ (fourth and fifth column, respectively) for all networks we consider in our numerical analysis. Computer-generated networks (specified in the second column)  are divided in non-self-similar, self-similar and perturbed self-similar (first column). The perturbation is made by rewiring a fraction $p=0.01$ of all links in the WS model and by adding a fraction $p=0.05$ or $p=0.01$ of all connections in the FM or AP networks, respectively. The third column specifies the type of transformation used to analyze the renormalization flow. We associate to each numerical value of $\nu$ and $x^*$ its error.}
\label{table}
\end{table}

\subsection{Effect of small perturbations on self-similar networks}
\label{sec:perssn}

\begin{figure*}
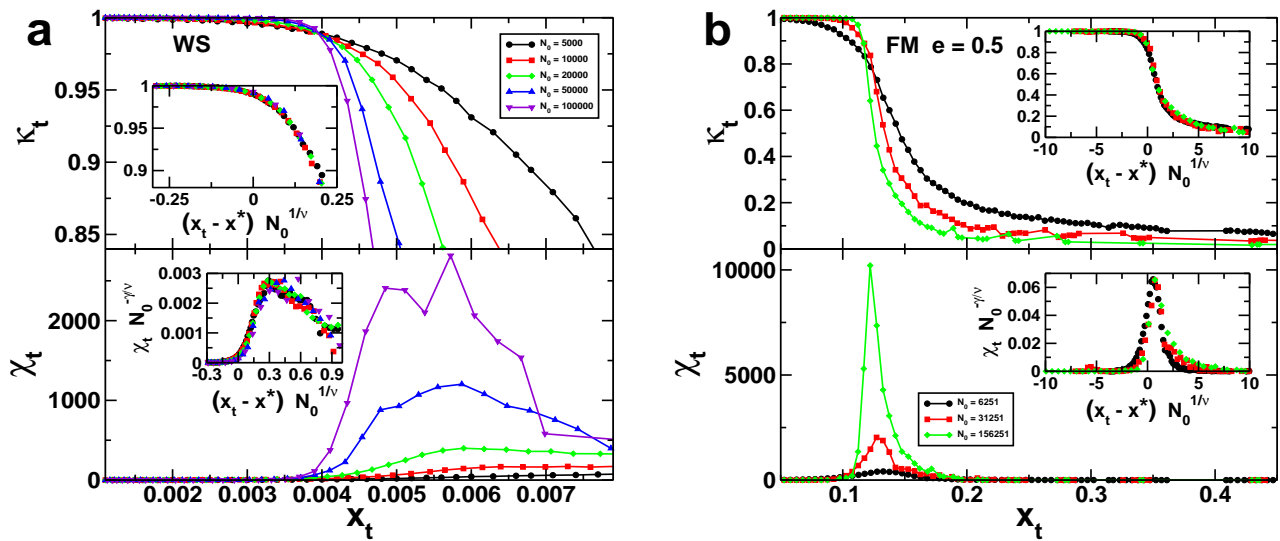

\includegraphics[width=0.45\textwidth, height=0.3\textheight]{fig9a}
\qquad
\includegraphics[width=0.45\textwidth, height=0.3\textheight]{fig9b}
\caption{Effect of a small random perturbation on
  renormalization flows.  The box covering has been performed by using
  RB with $r_B=1$.  (a) WS network with $\langle k \rangle =4$ and a
  fraction $p=0.01$ of randomly rewired links.  (b) FM network with
  $e=0.5$ and a fraction $p=0.05$ of added links.  The figures display
  $\kappa_t$ (a, b, top), and $\chi_t$ (a, b, bottom) as a function of
  the relative network size $x_t$. Comparing with
  Fig.~\ref{fig:fm}a, we see that the transformation yields a
  crossing of the $\kappa_t$-curves for  the FM
  networks. The crossing appears also for the WS networks, while in the 
  unperturbed case the $\kappa_t$-curves do not cross since these networks 
  correspond to linear chains (trivially self-similar). The exponents 
  are now very different from the unperturbed
  case: we recover $\nu=2$. The relation $\gamma=\nu$ seems to hold
  here as well.}
\label{fig:pert}
\end{figure*}

Self-similar objects correspond by definition to fixed points of the
transformation. To study the nature of these fixed points, we have
repeated the analysis of the renormalization flows for the
self-similar networks considered, but {\em perturbed} by a small
amount of randomness, through the addition or rewiring of a small
fraction $p$ of links.  The results are shown in Fig.~\ref{fig:pert}
for WS small-world networks, which are simply linear chains (trivially
self-similar) perturbed by a certain amount of
rewiring~\cite{watts98}, and FM networks with randomly added links. In
both cases we recover the behavior observed for non-self-similar
graphs, with a scaling exponent $\nu=2$ (this holds for all values of
$r_B$ or $\ell_B$ investigated, see also \cite{radicchi08}). This
clearly implies that the original self-similar fixed points are
unstable with respect to disorder in the connections, and highlights
once again the robustness of the exponent value $\nu=2$. Furthermore, the
statistical properties of the fixed point reached at $x^*$, when it
exists (i.e., for $r_B=1$ or $\ell_B=2$) are again the same as those
obtained starting from non-self-similar networks (see
Fig.~\ref{fig:fixpoint}). For these particular renormalization
flows (for $r_B=1$ or $\ell_B=2$), the picture obtained is
therefore a global flow towards the structure depicted in 
Fig.~\ref{fig:fixpoint}, with isolated unstable fixed points given by
the artificial self-similar graphs.

\section{Real networks}

\begin{figure*}
\begin{centering}
\includegraphics[width=0.45\textwidth, height=0.3\textheight]{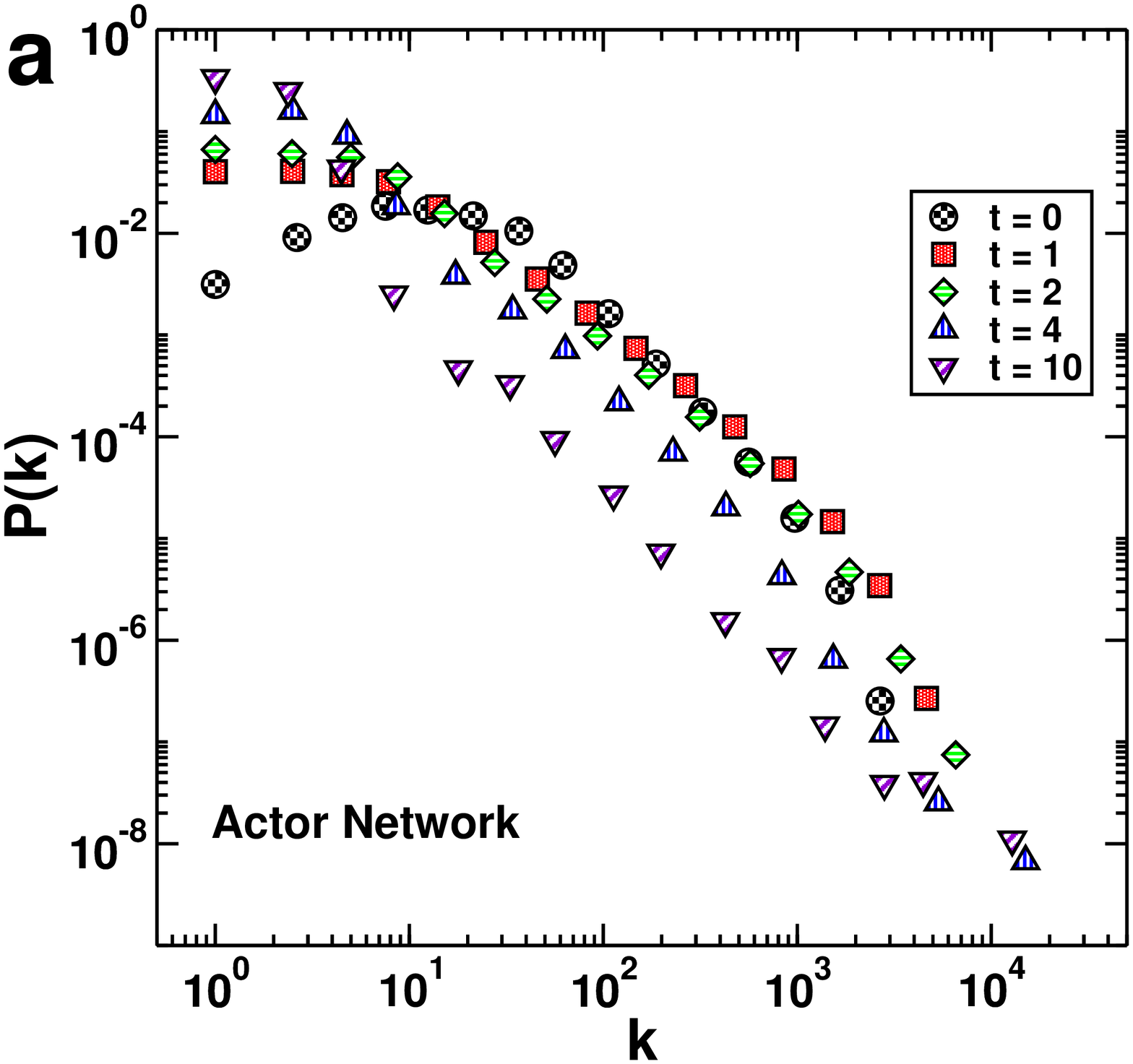}
\qquad
\includegraphics[width=0.45\textwidth, height=0.3\textheight]{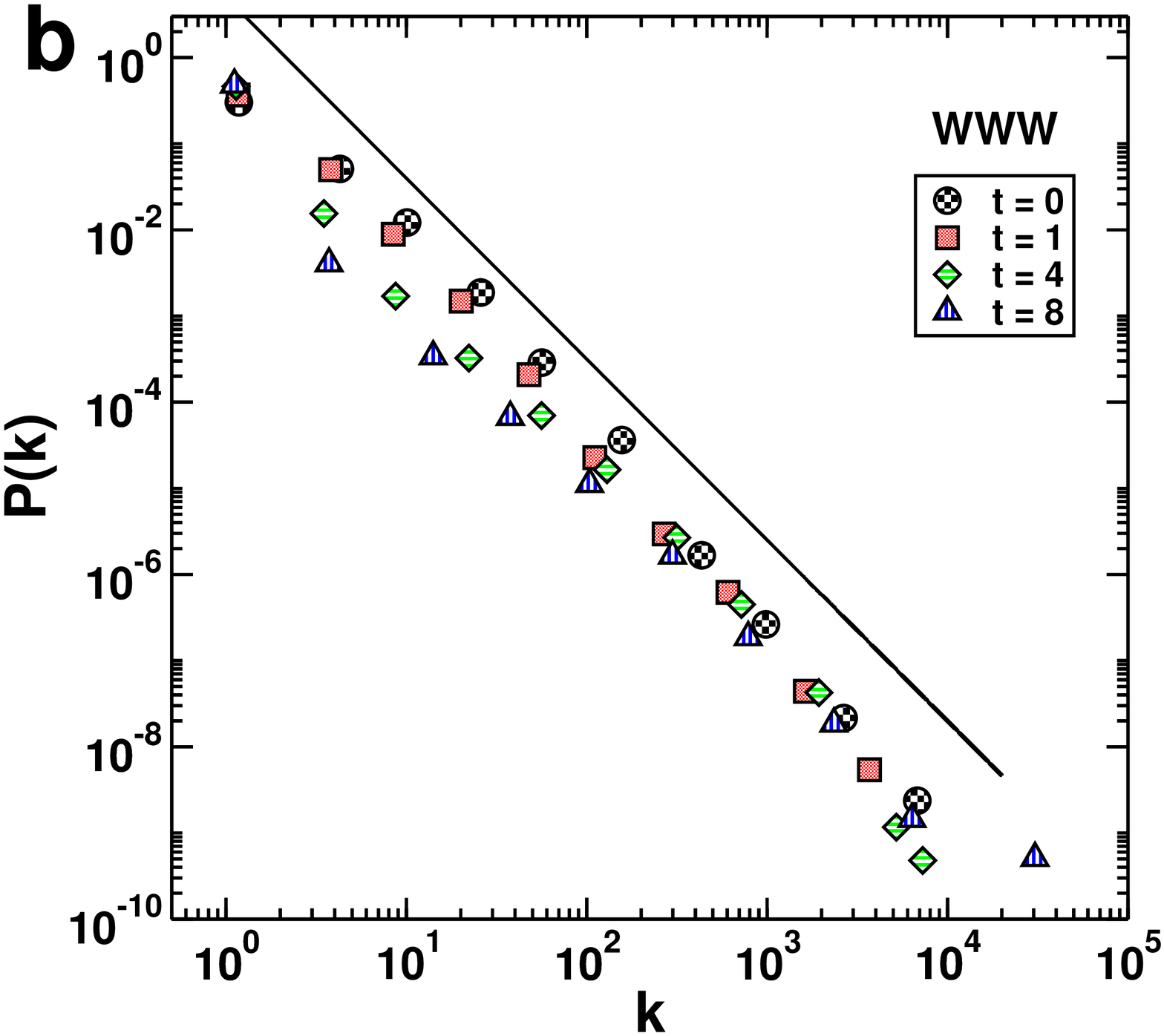}
\vskip .3cm
\includegraphics[width=0.45\textwidth, height=0.3\textheight]{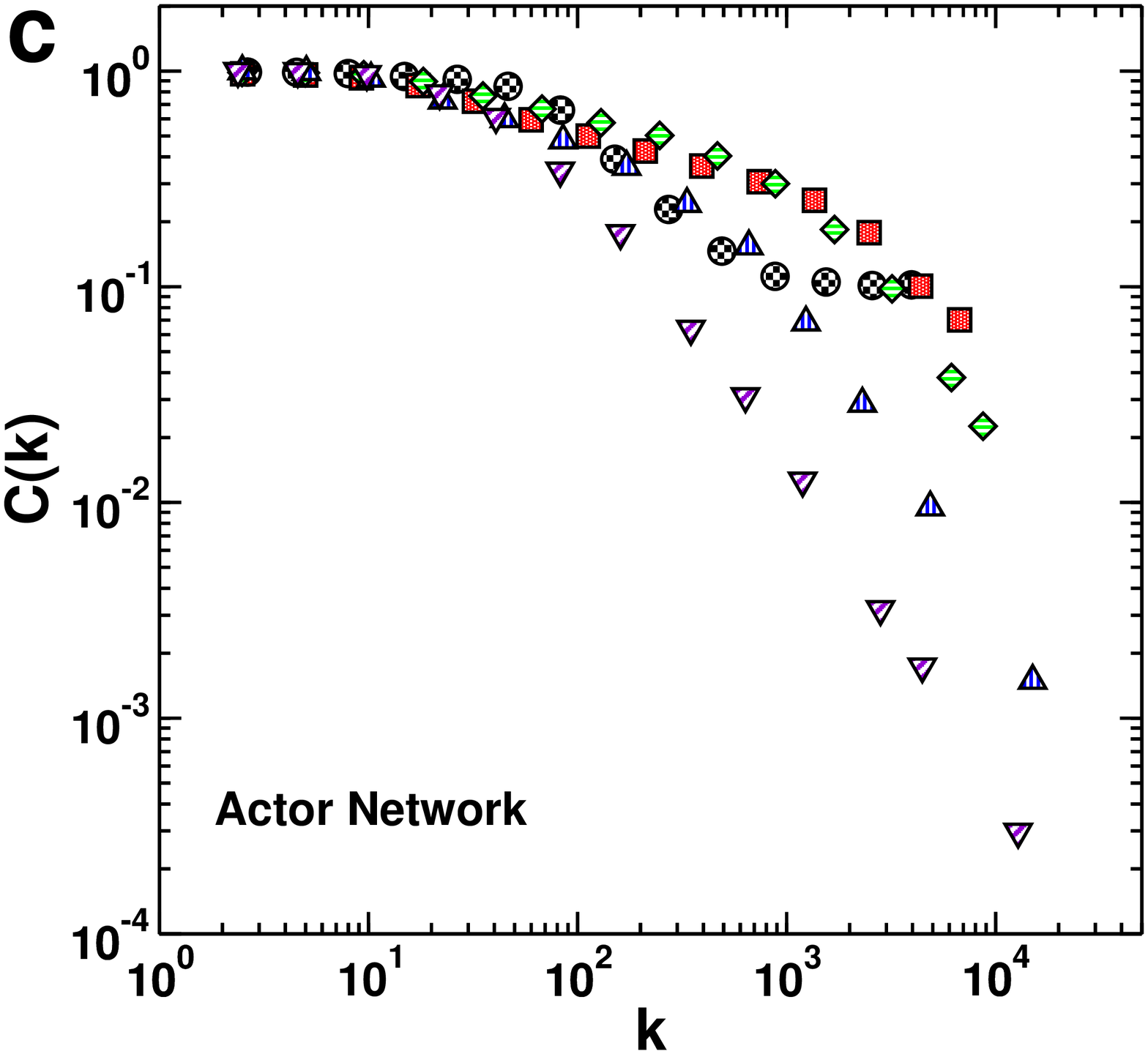}
\qquad
\includegraphics[width=0.45\textwidth, height=0.3\textheight]{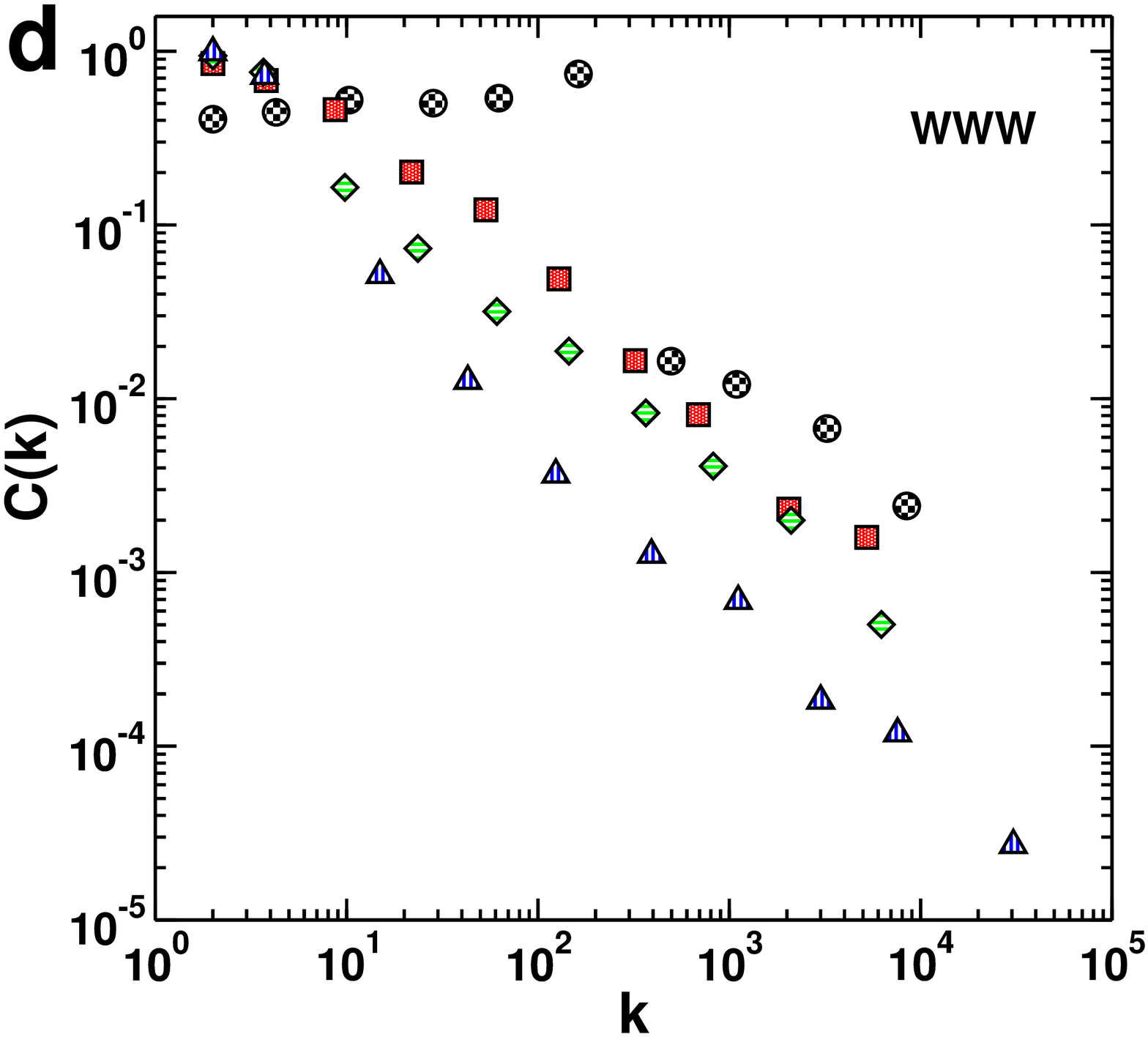}
\vskip .3cm
\includegraphics[width=0.45\textwidth, height=0.3\textheight]{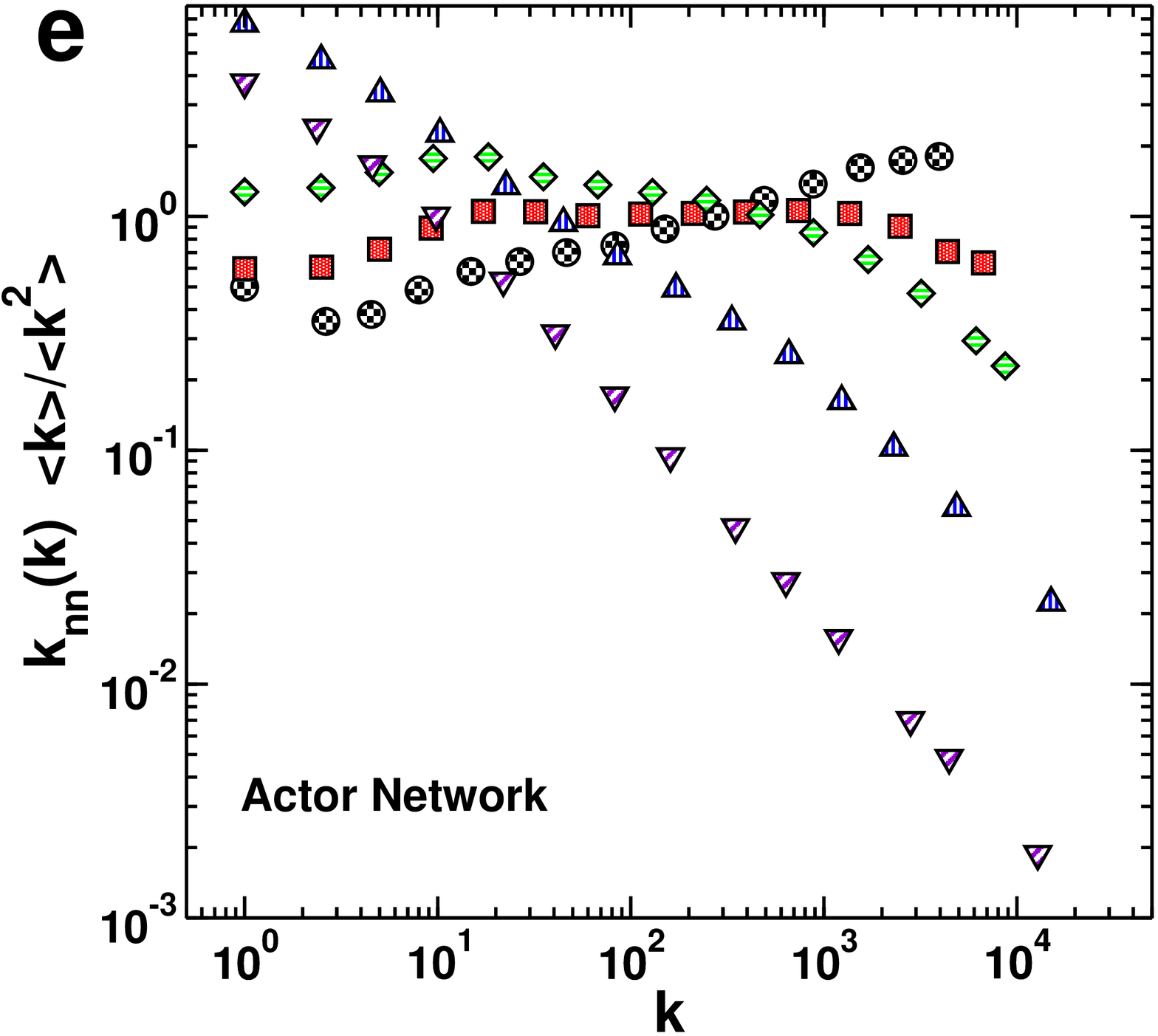}
\qquad
\includegraphics[width=0.45\textwidth, height=0.3\textheight]{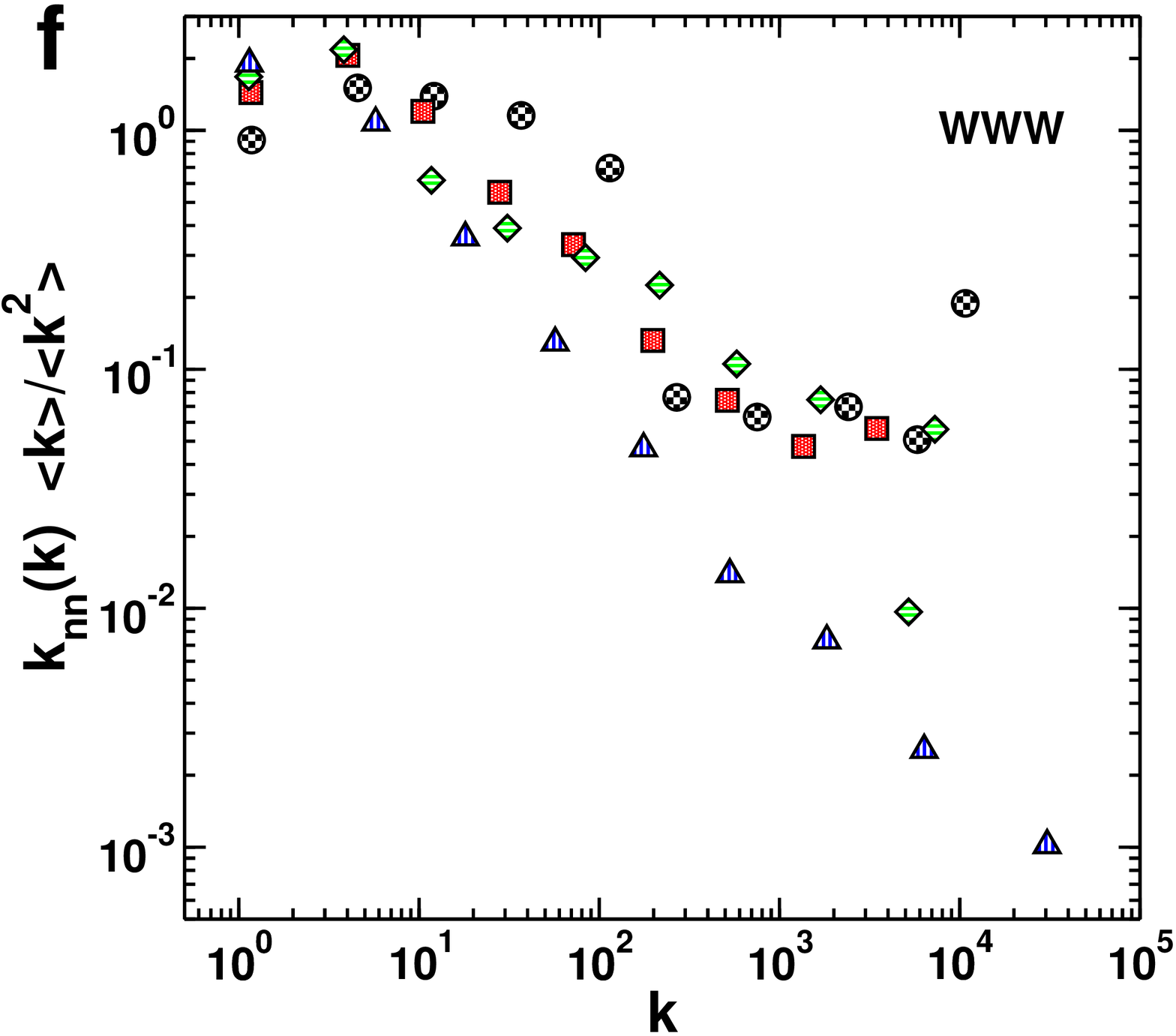}
\end{centering}
\caption{Statistical properties of real networks after
  $t$ steps of renormalization.  We consider two examples: a network
  of $392,340$ actors, where nodes are connected if the corresponding
  actors were cast together in at least one movie~\cite{barabasi99}
  (a, c and e); the link graph of the WWW, consisting of $325,729$ Web
  pages of the domain of the University of Notre Dame (Indiana, USA)
  and of their mutual hyperlinks~\cite{albert99} (b, d and f).  The
  box covering was performed with the GCA ($\ell_B=2$), but the
  results hold as well for other transformations. The clustering
  spectrum and the degree correlation pattern change drastically
  already after a single transformation. In particular, the actor
  network displays assortativity, but after two transformations it
  becomes disassortative.  The solid line in (b) has slope $-2.1$.}
\label{fig:real} 
\end{figure*}

For real-world networks, a finite-size scaling analysis is not
available because of the uniqueness of each sample. On the other hand,
it is still possible to apply repeatedly the renormalization
transformation and to study the evolution of the network properties (a similar numerical study has been performed also in~\cite{ichikawa08}). 
In Fig.~\ref{fig:real}, we measure some basic statistical properties
of two real networks along the renormalization flow. We consider the
Actor Network~\cite{barabasi99}, a graph constructed from the Internet
Movie Database~\footnote{www.imdb.com} where nodes are connected if
the corresponding actors were cast together in at least one movie, and
the link graph of the Web pages of the domain of the University of
Notre Dame (Indiana, USA)~\cite{albert99}. Both networks have been
claimed to be self-similar, since Eq.(\ref{eq:self}) holds for both of
them~\cite{song05}. On the one hand, the degree distributions $P(k)$
are only slightly affected by the renormalization transformation, and
retain their main characteristics even after several stages of
renormalization (in particular for the Web graph, see
Fig.~\ref{fig:real}b). This first result points towards an effective
self-similarity of $P(k)$ under the action of the renormalization
flow.  The degree distribution by itself is however not enough to
characterize complex networks, since many different topologies
can correspond to the same $P(k)$. Important information
is in particular encoded in the clustering coefficient spectrum
$C(k)$, defined as the average clustering coefficient of nodes of
degree $k$, and in the average degree of the neighbors of nodes of
degree $k$, $k_{nn}(k)$, which is a measure of the degree correlations between nearest neighbors in the graph.
In this context, Fig.~\ref{fig:real}c, d, e, and f show that even a single 
renormalization transformation induces large changes in 
these quantities. In this respect, the apparent self-similarity
exhibited by the degree distribution does not extend to higher
order correlation patterns.

\begin{figure*}[!ht]
\begin{center}
\includegraphics[width=0.45\textwidth, height=0.3\textheight]{fig11a}
\qquad
\includegraphics[width=0.45\textwidth, height=0.3\textheight]{fig11b}
\vskip .3cm 
\includegraphics[width=0.45\textwidth, height=0.3\textheight]{fig11c}
\qquad
\includegraphics[width=0.45\textwidth, height=0.3\textheight]{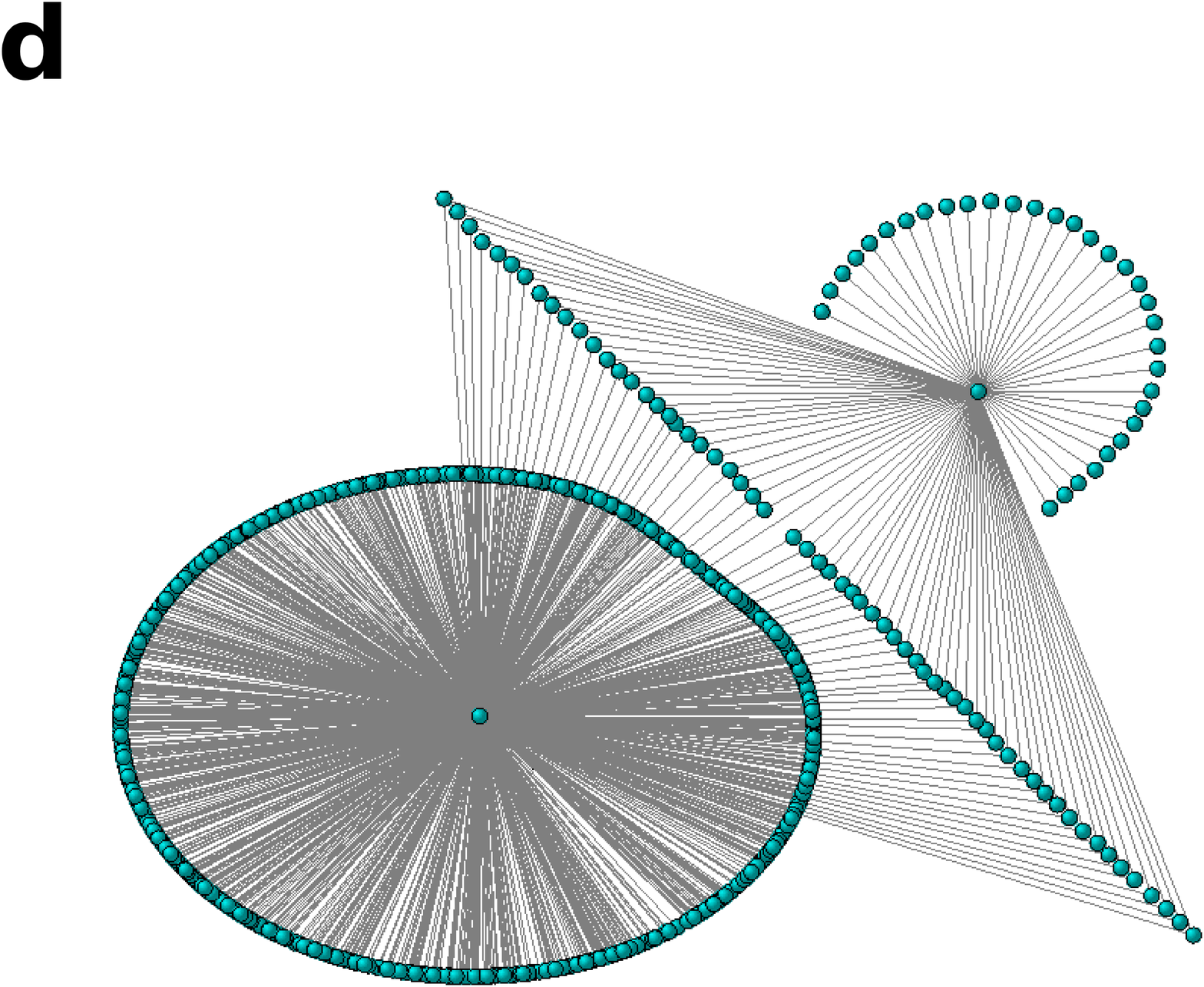}
\end{center}
\caption{Statistical properties of the 'fixed point' in
  the case of real networks. Renormalization has been performed by
  using GCA with $\ell_B=2$. The properties of the networks are
  measured after a certain number (less than ten) of renormalization
  steps. Dashed lines have the same slopes as those appearing in
  Fig.~\ref{fig:fixpoint}. The real networks considered in this figure
  are: the actor network~\cite{barabasi99}, the scientific
  collaboration network~\cite{newman01}, the network of Web pages of
  the domain of the University of Notre Dame~\cite{albert99} and the
  protein-protein interaction network of the yeast {\it Saccharomyces
    cerevisae}~\cite{colizza05}. (d) The graphical representation of
  the fixed point has been obtained by starting from the protein-protein
  interaction network of the yeast.}
\label{fig:fixpoint2}
\end{figure*}

Interestingly, in the case $r_B=1$ or $\ell_B=2$, the iteration of the
renormalization transformation leads all real networks investigated
[either self-similar or not, as defined by Eq.(\ref{eq:self})] towards
the same kind of structure (illustrated in Fig.~\ref{fig:fixpoint}) which is reached
by non-self-similar artificial networks (see
Fig.~\ref{fig:fixpoint2}). Note that, for real networks, no change in the
initial size can be performed, so we simply show the structure
obtained after a few steps of renormalization, which remains
stable for many steps due to the peculiarity of the case $r_B=1$ or $\ell_B=2$,
as explained above.

All these results allow us to discuss the exact self-similarity of real-world networks: as we have seen in the case of
computer-generated self-similar networks, all fixed-points correspond
to strongly regular topologies and minimal perturbations are enough to
break the picture of self-similarity. Since randomness is an
unavoidable element in real complex networks, exact
self-similarity should not be observed in them. 
The randomness of their topology is amplified when
renormalization is iterated. Real-world networks which are
self-similar according to Eq.(\ref{eq:self}) could however a priori be
arbitrarily close to an actual fixed point of the
renormalization. This actually raises the important issue of
defining and measuring a distance in the space of networks. However Fig.~\ref{fig:real} shows that few renormalization steps
(often a single one) are enough to substantially modify the network
structure in our examples, so that it seems to us hard to sustain that they are close
to a fixed point.


\label{sec:real}

\section{Summary and conclusions}
\label{sec:concl}
In this paper we have presented a detailed analysis of the method,
introduced in~\cite{radicchi08}, to study renormalization flows
of complex networks.  The method is applied using the two most popular
techniques for network renormalization: greedy coloring~\cite{song06}
and random burning~\cite{goh06}. Independently of the
algorithm, we have shown that a simple scaling rule [i.e.,
  Eq.(\ref{eq:scaling})] is able to describe the renormalization flow
of any computer-generated network. A single scaling exponent $\nu$ is
needed in order to classify networks in universality classes: all
non-self-similar networks belong to the same universality class
characterized by $\nu=2$; self-similar networks, on the
other hand, belong to other universality classes, generally
identified by values of the scaling exponent $\nu$ different from
$2$. Self-similar networks represent by definition fixed points
of the renormalization transformation, since the statistical
properties of these networks remain invariant
after renormalization. They are actually unstable fixed points of the
renormalization transformation: minimal random perturbations are
indeed able to lead the flow out of these fixed points.
The numerical study presented here confirms and extends the validity of the results already anticipated in~\cite{radicchi08}.

In addition, we have performed an analysis of the effect of the iterated renormalization transformation on real networks. Unfortunately, the same technique, introduced in~\cite{radicchi08}, cannot be directly applied to real networks. In
this case, in fact, a finite-size scaling analysis cannot be performed
since any real network has a fixed size. Nevertheless, the usual simple
measures of the network structure, taken after a few iterations of
renormalization, reveal that the transformation modifies the
topological properties of the network. Furthermore, the repeated
renormalization of real networks produces flows converging to the same fixed point
structure, when it exists, as the one found in the case of
computer-generated non-self-similar networks.

\end{document}